\documentclass[fleqn,twoside]{article}
\usepackage{espcrc2}

\newcommand{\be}{\begin{equation}}
\newcommand{\ee}{\end{equation}}
\newcommand{\bem}{\begin{displaymath}}
\newcommand{\eem}{\end{displaymath}}
\newcommand{\ba}{\begin{eqnarray}}
\newcommand{\ea}{\end{eqnarray}}
\newcommand{\re}[1]{(\ref{#1})}

\newcommand{\1}{^{-1}}

\newcommand{\dg}{^{\dagger}}

\newcommand{\e}{\mbox{e}}

\newcommand{\f}{{\mbox{\scriptsize f}}} 
\newcommand{\g}{{\mbox{\scriptsize g}}} 
\newcommand{\bg}{\bar{g}} 
\newcommand{\bG}{\bar{G}} 
\newcommand{\G}{{\cal B}} 
\newcommand{\ga}{\gamma_5}
\newcommand{\h}{\frac{1}{2}}
\newcommand{\Id}{\mbox{1\hspace{-1.05mm}l}}

\newcommand{\mo}[1]{^{(\mbox{\scriptsize #1})}}
\newcommand{\N}{{\cal N}}
\newcommand{\Op}{{\cal O}} 
\newcommand{\Pa}{{\cal P}} 
\newcommand{\bP}{\bar{P}}

\newcommand{\T}{{\cal T}} 
\newcommand{\Tr}{\mbox{Tr}} 
\newcommand{\bu}{\bar{u}} 
\newcommand{\U}{{\cal U}} 
\newcommand{\vp}{\varphi} 
\newcommand{\bw}{\bar{w}}
\newcommand{\W}{{\cal W}} 

\newcommand{\AmS}{{\protect\the\textfont2
  A\kern-.1667em\lower.5ex\hbox{M}\kern-.125emS}}

\title{\vspace*{-10mm}
\raisebox{0.8cm}[0pt][0pt]{\makebox[0pt][l]{\parbox{16cm}{\normalsize%
\mbox{}\hfill HUB-EP-03/52}}}\\
General chiral gauge theories\thanks{
Talk presented at Lattice 2003, Tsukuba, Japan.}}

\author{Werner Kerler \address{Institut f\"ur Physik, 
        Humboldt-Universit\"at, D-12489 Berlin, Germany}%
}
       
\begin{document}

\begin{abstract}
Only requiring that Dirac operators decribing massless fermions on the 
lattice decompose into Weyl operators we arrive at a large class of them. 
After deriving general relations from spectral representations we study 
correlation functions of Weyl fermions for any value of the index, stressing
the related conditions for basis transformations and getting the precise 
behaviors under gauge and CP transformations. Using the detailed structure 
of the chiral projections we also obtain a form of the correlation functions 
with a determinant in the general case.
\vspace{1pc}
\end{abstract}

\maketitle
\thispagestyle{empty}

\section{INTRODUCTION}

Reconsidering chiral gauge theories on the lattice we generalize the basic 
structure which has been introduced in the overlap formalism of Narayanan 
and Neuberger \cite{na93} and in the formulation of L\"uscher \cite{lu98}. 
Only requiring that the Dirac operator $D$ allows a decomposition into Weyl 
operators we still extend the large class of operators decribing massless 
fermions on the lattice which we have found recently \cite{ke02}. In addition 
to Ginsparg-Wilson (GW) fermions \cite{gi82} this class includes the ones 
proposed by Fujikawa \cite{fu00} and the extension of the latter \cite{fu02}.

Noting that the operators $D$ are functions of a basic unitary and 
$\ga$-Hermitian operator $V$ we introduce chiral projections of form 
\cite{ke03}
\be
P_{\pm}=\h\big(\Id\pm\ga G\big),\quad\bP_{\pm}=\h\big(\Id\pm \bG\ga\big),
\label{PR} 
\ee
with functions $G(V)$ and $\bG(V)$ satisfying
\be
G\1=G\dg=\ga G\ga,\quad\bG\1=\bG\dg=\ga\bG\ga.
\ee
Requiring the decomposition
\be
D=\bP_+DP_-+\bP_-DP_+
\ee
then leads to the relations 
\be
\bP_{\pm}DP_{\mp}= DP_{\mp}=\bP_{\pm}D
\ee 
and gives the general condition 
\be
D+D\dg\bG G=0.
\label{DGG} 
\ee
The forms considered in Refs.~\cite{lu98} and \cite{ha02} in the GW case 
correspond to the special choices $G=V$, $\bG=\Id$ and $G=((1-s)\Id+sV)/\N$,
$\bG=(s\Id+(1-s)V)/\N$ with a real parameter $s$ and normalization factor
$\N$, respectively. 

\section{GENERAL RELATIONS}

We start from the spectral representation of $V$
\ba
V=P_1^{(+)}+P_1^{(-)}-P_2^{(+)}-P_2^{(-)}\quad\nonumber\\+\sum_{k\;(0<\vp_k
<\pi)}\big(\e^{i\vp_k} P_k\mo{I}+\e^{-i\vp_k} P_k\mo{II}\big),
\ea
where the orthogonal projections satisfy
$\ga P_j^{(\pm)}=P_j^{(\pm)}\ga=\pm P_j^{(\pm)}$, 
$\ga P_k\mo{I}=P_k\mo{II}\ga$,
which implies that one has 
\be 
N_+(1)-N_-(1)+N_+(-1)-N_-(-1)=0,
\label{SU}
\ee
\be 
N_{+}(1)-N_{-}(1)=\h\mbox{Tr}(\ga V)
\label{IV}
\ee
for $N_{\pm}(1)=\mbox{Tr}\big(P_1^{(\pm)}\big)$, 
$N_{\pm}(-1)=\mbox{Tr}\big(P_2^{(\pm)}\big)$, and 
\be
\mbox{Tr}\big(\ga P_k\mo{I}\big)=\mbox{Tr}\big(\ga P_k\mo{II}\big)=0,
\ee
\be
\mbox{Tr}\,P_k\mo{I}=\mbox{Tr}\,P_k\mo{II}.
\ee

The representation of $D=F(V)$ then becomes
\bem
D=f(1)\big(P_1^{(+)}+P_1^{(-)}\big)+f(-1)\big(P_2^{(+)}+P_2^{(-)}\big)
\eem
\be
+\sum_{k\;(0<\vp_k<\pi)}\big(f(\e^{i\vp_k})P_k\mo{I}+f(\e^{-i\vp_k})
P_k\mo{II}\big),
\label{specd}
\ee
with the conditions on the spectral functions
\be
f(1)=0,\quad f(-1)\ne0,\quad f^*(v)=f(v^*),
\ee
corresponding to masslessness, allowance for a non-zero index, 
$\ga$-Hermiticity, respectively. The index of $D$ then gets $I=N_+(1)-N_-(1)$,
which thus is given by \re{IV} and subject to \re{SU}.

For $G$ and $\bG$ analogous representations to \re{specd} hold with spectal
functions $g$ and $\bg$, respectively. The latter satisfy $|g|^2=1$, 
$g^*(v)=g(v^*)$, $|\bg|^2=1$, $\bg^*(v)=\bg(v^*)$. In terms of spectral 
functions the general condition \re{DGG} now reads
\be
f+f^*\bg g=0.
\ee
Because of $f(-1)\ne0$ this implies
\be
\bg(-1)=-g(-1),
\label{bg-}
\ee
with the important consequence that $\bG$ and $G$ must be generally different.

To have $I=\bar{N}-N$ for the index, where $\bar{N}=\Tr\,\bP_+$, $N=\Tr\,P_-$, 
one needs $\bg(1)=g(1)=1$. Then for the choices $g(-1)=-\bg(-1)=\pm1$ one
gets $N=\h\Tr\,\Id-I$, $\bar{N}=\h\Tr\,\Id$ and $N=\h\Tr\,\Id$, $\bar{N}=\h
\Tr\,\Id+I$, respectively.

The spectral functions have been used in Ref.~\cite{ke02} to construct various
concrete examples of Dirac operators. The respective methods have been seen
in Ref.~\cite{ke03} to extend to the more general class of operators 
satisfying \re{DGG}.

\section{CORRELATION FUNCTIONS}

General fermionic correlation functions can be written as \cite{ke03}
\be
\langle\psi_{\sigma_{r+1}}\ldots\psi_{\sigma_N}\bar{\psi}_{\bar{\sigma}_{r+1}}
\ldots\bar{\psi}_{\bar{\sigma}_{\bar{N}}}\rangle_{\f}
\label{COR}
\ee
\bem
=\frac{1}{r!}\sum_{\bar{\sigma}_1\ldots\bar{\sigma}_r}\sum_{\sigma_1,\ldots,
\sigma_r}\bar{\Upsilon}_{\bar{\sigma}_1\ldots\bar{\sigma}_{\bar{N}}}^*
\Upsilon_{\sigma_1\ldots\sigma_N}D_{\bar{\sigma}_1\sigma_1}\ldots
D_{\bar{\sigma}_r\sigma_r}
\eem
with alternating multilinear forms $\Upsilon_{\sigma_1\ldots\sigma_N}$ and 
$\bar{\Upsilon}_{\bar{\sigma}_1\ldots\bar{\sigma}_{\bar{N}}}$, which are
explicitly represented by
\be
\Upsilon_{\sigma_1\ldots\sigma_N}=\sum_{j_1,\ldots,j_N=1}^N\epsilon_{j_1, 
\ldots,j_N}u_{\sigma_{1}j_{1}}\ldots u_{\sigma_Nj_N}
\label{FO}
\ee
and an analogous expression for 
$\bar{\Upsilon}_{\bar{\sigma}_1\ldots\bar{\sigma}_{\bar{N}}}$. The bases in
such expressions satisfy 
\be
P_-=uu\dg,\; u\dg u=\Id_{\rm w},\;\bP_+=\bu\bu\dg,\;\bu\dg\bu=\Id_{\rm \bw},
\label{uu}
\ee 
where $\Id_{\rm w}$ and $\Id_{\rm \bw}$ are the identity operators in the
spaces of the Weyl degrees of freedom. Comparing with vector theory it is
seen that instead of $\epsilon_{\sigma_1\ldots \sigma_K}$ and $\epsilon_{
\bar{\sigma}_1\ldots\bar{\sigma}_K}$ with $K=\Tr\,\Id$ 
there, one has $\Upsilon_{\sigma_1\ldots\sigma_N}$ and  
$\bar{\Upsilon}_{\bar{\sigma}_1\ldots\bar{\sigma}_{\bar{N}}}$ here. 

By \re{uu} the bases are fixed up to unitary transformations, $u^{(S)}=uS$, 
$\bu^{(\bar{S})}=\bu\bar{S}$. In addition requiring unimodularity,  
${\det}_{\rm w}S=1$, ${\det}_{\rm\bw}\bar{S}=1$, the correlation functions
\re{COR} get invariant. While without this additional restriction the 
transformations $S$ connect all bases of the subspace on which $P_-$ projects,
the unimodular $S$ connect only subsets thereof. The total set of bases 
$u^{(S)}$ thus decomposes into subsets. Because the formulation of the theory 
has to be restricted to one of such subsets (which are not equivalent)
there is the question to which one of them. Analogous considerations
apply to the bases $\bu^{(\bar{S})}$. 

\section{GAUGE TRANSFORMATIONS}

Considering the gauge transformation $P_-'=\T P_-\T\dg$ for $G\ne\Id$ with 
$[\T,P_-]\ne0$, given a solution $u$ satisfying \re{uu} then $u'=\T u S$ is 
a solution of the transformed conditions $P_-'=u'u'\,\!\dg$ and $u'\,\!\dg u'
=\Id_{\rm w}$. With the restriction to unimodular $S$ this constitutes a
mapping between the respective subset and the transformed subset of bases.
With analogous considerations for $\bu$ it becomes obvious that for $G\ne\Id$,
$\bG\ne\Id$ the correlation functions \re{COR} transform gauge-covariantly,
\be
\langle\psi_{\sigma_1'}'\ldots\psi_{\sigma_L'}'\bar{\psi}_{\bar{\sigma}_1'}'
\ldots\bar{\psi}_{\bar{\sigma}_{\bar{L}}'}'\rangle_{\f}'
\label{COV}
\ee
\bem
=\sum_{\sigma_1,\ldots,\sigma_L}\sum_{\bar{\sigma}_1,\ldots,\bar{\sigma}_L}
\T_{\sigma_1'\sigma_1}\ldots\T_{\sigma_L'\sigma_L}
\eem
\bem
\qquad\langle\psi_{\sigma_1}\ldots\psi_{\sigma_L}\bar{\psi}_{\bar{\sigma}_1}
\ldots\bar{\psi}_{\bar{\sigma}_{\bar{L}}}\rangle_{\f}\,
\T_{\bar{\sigma}_1\bar{\sigma}_1'}\dg\ldots
\T_{\bar{\sigma}_{\bar{L}}\bar{\sigma}_{\bar{L}}'}\dg.
\eem

For $\bG=\Id$ with $[\T,\bP_+]=0$ and $\bP_+'=\bP_+$, given a gauge-field
independent solution $\bu_{\rm c}$ of \re{uu}, also $\bu=\bu_{\rm c} S$ with 
unimodular $S$ is a solution, representing the respective invariant subset of
bases. This solution can be rewritten as $\bu=\T\bu_{\rm c}\bar{S}_{\T}\dg S$
with $\bar{S}_{\T}=\bu_{\rm c}\dg\T\bu_{\rm c}$ (which is unitary, however,
in general not unimodular). With this for the correlation functions in the 
case $G\ne\Id$, $\bG=\Id$ one gets the behavior \re{COV} multiplied by the 
constant phase factor ${\det}_{\rm\bw} \bar{S}_{\T}$. Evaluation of this factor
\cite{ke03} gives ${\det}_{\rm\bw}\bar{S}_{\T}=\e^{\h{\rm Tr}\,\G}$ where 
$\T=\e^{\G}$ and $\G_{n'n}=i\delta^4_{n'n}\sum_{\ell}b_n^{\ell}T^{\ell}$.
Thus the additional condition $\mbox{tr}_{\g} T^{\ell}=0$ on the generators 
$T^{\ell}$ is needed to get rid of that factor. 

\section{CP TRANSFORMATIONS}

With the charge conjugation matrix $C$, $\Pa_{n'n}=\delta^4_{n'\tilde{n}}$, 
$U_{4n}^{\rm CP}=U_{4\tilde{n}}^*$, $U_{kn}^{\rm CP}=U_{k,\tilde{n}- 
\hat{k}}^*$ for $k=1,2,3$, where $\tilde{n}=(-\vec{n},n_4)$, we have
\be
\Op(\U^{\rm CP})=\W\big(\Op(\U)\big)^{\rm T}\W\dg,\quad\W=\Pa\gamma_4C\dg
\ee
for $V$, $D$, $G$, $\bar{G}$, while for the
projections
$P_-^{\rm CP}(\U^{\rm CP})$, $\bP_+(\U)$,
$\bP_+^{\rm CP}(\U^{\rm CP})$, $P_-(\U)$, we get
\be
P_-^{\rm CP}=\W\bP_+^{\rm T}\W\dg,\quad \bP_+^{\rm CP}=\W P_-^{\rm T}\W\dg,
\ee
\be
P_-^{\rm CP}=\h\big(\Id-\ga\bG\big),\quad\bP_+^{\rm CP}=\h\big(\Id+G\ga\big).
\label{PRC}
\ee
It is seen that \re{PRC} differs from \re{PR} by an interchange of $G$ and 
$\bG$. Since in \re{DGG} only the product enters, the interchanged choice
is associated to the same $D$. Because according to \re{bg-} one has generally
$\bG\ne G$, one cannot get the symmetric situation of continuum theory.

Given  solutions $u$ and $\bu$ of \re{uu}, then $u^{\rm CP}=\W\bu^*\bar{S}$
and $\bu^{\rm CP}=\W u^*S$ are solutions of the CP transformed conditions.
With unimodular $S$ and $\bar{S}$ this leads to the transformations
\be
\langle\psi_{\sigma_1'}^{\rm CP}\ldots\psi_{\sigma_L'}^{\rm CP}\bar{\psi}_{
\bar{\sigma}_1'}^{\rm CP}\ldots\bar{\psi}_{\bar{\sigma}_{\bar{L}}'}^{\rm CP}
\rangle_{\f}^{\rm CP}
\ee
\bem
=\sum_{\sigma_1,\ldots,\sigma_L}\sum_{\bar{\sigma}_1,\ldots,\bar{\sigma}_{
\bar{L}}} \W_{\bar{\sigma}_1
\bar{\sigma}_1'}\dg\ldots\W_{\bar{\sigma}_{\bar{L}}\bar{\sigma}_{\bar{L}}'}\dg
\eem
\bem
\quad\langle\psi_{\bar{\sigma}_1}\ldots\psi_{\bar{\sigma}_{\bar{L}}}
\bar{\psi}_{\sigma_1}\ldots\bar{\psi}_{\sigma_L}
\rangle_{\f},\W_{\sigma_1'\sigma_1}\ldots\W_{\sigma_L'\sigma_L}.
\eem

In Ref.~\cite{fu02} the special form of Ref.~\cite{ha02} for $G$, $\bG$ has
been used together with more general $D$. A singularity has been encountered 
if a symmetric situation for CP properties has been enforced, which here has 
been seen to be generally excluded by $\bG\ne G$. The exchange of parameters 
under CP transformations there corresponds to the interchange of $G$ and $\bG$
in the general case here. 

\section{FORM WITH DETERMINANT}

Using the relations of Sections 1 and 2 and choosing $g(-1)=-\bg(-1)=\pm1$
we get for the chiral projections the representations \cite{ke03} 
\ba
P_-=P_1^{(-)}+P_2^{(\mp)}+\sum_kP_k^{[-]},\\
\bP_+=P_1^{(+)}+P_2^{(\mp)}+\sum_k\bP_k^{[+]},
\ea
in which the newly introduced projections satisfy
\be
\Tr\,P_k^{[-]}=\Tr\,\bP_k^{[+]}=\Tr\,P_k\mo{I}=\Tr\,P_k\mo{II},
\ee
\be
D\dg\bP_k^{[+]}D=|f(\e^{i\vp_k})|^2P_k^{[-]}.
\label{DPD}
\ee
We thus have $N-N_-(1)=\bar{N}-N_+(1)=\tilde{N}$, so that removing zero
modes from $\bu\dg D u$ the reduced matrix $\tilde{M}$ gets quadratic. 
Starting from basis representations
of the $P_k^{[-]}$ and using \re{DPD} one gets $\e^{i\Theta_k}|f(\e^{i\vp_k})|$
for the diagonal elements and zero otherwise. For $P_2^{(\mp)}$ one remains
with the diagonal elements $\e^{i\Theta}|f(-1)|$. The phases $\Theta_k$ and
$\Theta$ can be freely choosen. The reduced chiral determinant ${\det}_{
\tilde{N}}\tilde{M}$ thus becomes
\be
\big(\e^{i\Theta}|f(-1)|\big)^{N_{\mp}(-1)}
\prod_k\big(\e^{i\Theta_k}|f(\e^{i\vp_k})|\big)^{N_k}
\ee
\vspace*{-2mm}and for the correlation functions \re{COR} we have \cite{ke03}  
\bem
\sum_{\sigma_{r+1}',\ldots,\sigma_N'}\epsilon\,_{\sigma_{r+1}
\ldots\sigma_N}^{\sigma_{r+1}'\ldots\sigma_N'}\;\sum_{\bar{\sigma}_{r+1}',
\ldots,\bar{\sigma}_{\bar{N}}'}\epsilon\,_{\bar{\sigma}_{r+1}\ldots
\bar{\sigma}_{\bar{N}}}^{\bar{\sigma}_{r+1}'\ldots\bar{\sigma}_{\bar{N}}'}\;
\eem
\bem
{\textstyle\frac{1}{(\tilde{N}-r)!}}(\tilde{P}_-\tilde{D}\1\tilde{\bP}_+)_{\sigma_{r+1}'
\bar{\sigma}_{r+1}'}\ldots(\tilde{P}_-\tilde{D}\1
\tilde{\bP}_+)_{\sigma_{\tilde{N}}'\bar{\sigma}_{\tilde{N}}'}\;
\eem
\bem
u_{\sigma_{\tilde{N}+1},\tilde{N}+1}\ldots u_{\sigma_NN}\;
\bu_{{\tilde{N}}+1,\bar{\sigma}_{{\tilde{N}}+1}}\dg\ldots
\bu_{\bar{N}\bar{\sigma}_{\bar{N}}}\dg\;{\det}_{\tilde{N}}\tilde{M}
\eem
where the operators $\tilde{P}_-$, 
$\tilde{D}$, $\tilde{\bP}_+$ are the ones restricted to the subspace without 
zero modes. 

\vspace*{2mm}
I wish to thank Michael M\"uller-Preussker and his group for their kind
hospitality.

\end{document}